\documentclass[a4paper,twocolumn]{esapub2005}
\usepackage{times}
\usepackage{natbib}

\usepackage{graphicx}
\usepackage{epsfig}

\def\INT{{\it INTEGRAL }}
\def\SS{{SS 433 }}

\title{INTEGRAL Observations of SS433: Analysis of Precessional and Orbital 
X-ray Periodicities}

\author{A.M. Cherepashchuk}
\affil{
Sternberg Astronomical Institute, Universitetsky pr. 13, 119992, Moscow, Russia
} 
\author{R.A. Sunyaev}
\affil{  Space Research Institute, Russian Academy of Sciences,
              Profsoyuznaya 84/32, 117810 Moscow, Russia
}
\affil{
 Max-Planck-Institute f\"ur Astrophysik,
              Karl-Schwarzschild-Str. 1, D-85740 Garching bei M\"unchen,
              Germany,
}

\author{E.V. Seifina}
\author{   E.A. Antokhina}
\author{   D.I. Kosenko}
\affil{
Sternberg Astronomical Institute, Universitetsky pr. 13, 119992, Moscow, Russia
}
\author{   S.V. Molkov}
\affil{  Space Research Institute, Russian Academy of Sciences,
              Profsoyuznaya 84/32, 117810 Moscow, Russia
}
\author{   N.I. Shakura}
\author{   K.A. Postnov}
\author{   A.N. Timokhin}
\author{   I.E. Panchenko}
\affil{
Sternberg Astronomical Institute, Universitetsky pr. 13, 119992, Moscow, Russia
}
\begin{document}
\maketitle

\abstract{
Hard X-ray \INT observations of \SS carried out during 2003-2005 years are presented. Analysis of precessional 
and orbital variability is presented. The width of X-ray eclipse in the $25-50$~keV range at the precessional phase $\psi=0.1$
(accretion disk is open to observer) is higher than that in the Ginga $18.4-27.6$~keV range. This fact 
suggests existance the presence of hot extended corona around the supercritical accretion disk.
Spectrum of hard X-rays in the range $10-200$~keV does not change with the precessional phase which also suggests that 
hard X-ray flux is generated in the hot extended corona around the accretion disk. 
The parameters of this hot corona are: $kT=23-25$~keV, $\tau = 1.8-2.8$. Mass ratio estimated  from the analysis of the ingress 
part of the eclipse light curve is in the range $q=m_x/m_v=0.3-0.5$.
}

\section{Introduction}
\SS is a massive eclipsing X-ray binary with precessing supercritical accretion disk and relativistic jets.
Narrow  collimated relativistic jets ($v=0.26c$) are precessing with the period $P_{prec}=162^d.5$. 
Orbital period is $13^d.08$ (\cite{Gor98}).
There are problems with optical classification of the determination of its radial velocity curve 
(\cite{Gies02}, \cite{Hillwig04}, \cite{Cher04}, \cite{Cher05}, \cite{Barnes05}).
This unique X-ray binary at an advanced evolutionary stage has been investigated in optical, radio and X-ray 
ranges (see review of Fabrika \cite{Fab04} and references therein).

First \INT observations of \SS in hard X-rays gave a surprise: \SS is a hard X-ray source with emission clearly
detected up to $100$~keV. We concluded that \SS is a supercritical microquasar with hard X-ray spectrum
(\cite{Cher03}). \SS  was observed by \INT in A01-A03 from 2003 to 2005.

\section{Observations and data reduction}

In this work all publically avaliable data on \SS obtained by \INT 
from March, 2003 to November, 2004 (AO-1,2) and the results of our pointed
observations at October 12-22, 2005 (AO-3) are presented.

The data reduction was performed by the Offline Science Analysis (OSA) software
version 5.1, developed by the \INT science data center (ISDC, http://isdc.unige.ch 
\cite{Cour03}).

The light curves where  reconstructed  by processing all the available observations,
obtained by the ISGRI/IBIS detector in the $25-50$~keV energy band. The light curve
points where taken from the mosaic images integrated by each 10 ScW (Science Windows) with a 
standard background approximation procedure. Thus, a typical exposition time
for each light curve point is $\sim 20$~ksec.

The variability analysis is based on the precessional and orbital ephemeris of \SS
by \cite{Fab04}. The orbital minima:
$$
    JD_{Min I He}= 2450023.62+13.08211*E
$$
and the maximal emission lines separation $T_3$ moments: 
$$
    JD_{T3}=2443507.47+162.375*E1,
$$
where $E$ and $E1$ are the orbital and precession cycle numbers respectively.

\begin{figure}[ht]
\epsfig{file=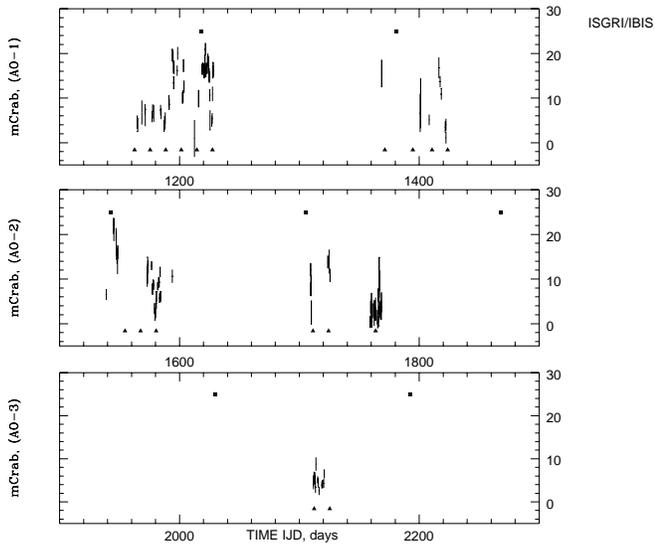,width=\linewidth}
\caption{Observations of \SS in $2003-2005$. The squares indicate the precessional face-on ($T_3$) moments, and the triangles indicate
the orbital minima. The units of the horizontal axis are $IJD=JD-2451544.5$.
}\label{obsvstime}
\end{figure}

During the 3 years of \INT observations (2003-2005) cover a wide range
of the precession phases of \SS. The three observation sets of AO-1
are performed at $\psi=0.72-0.88$ (incontiunosly), $0.95-0.1$ (continuosly); 
AO-2: $\psi= 0.0-0.15$ (incontinuously), $0.22-0.23$ (continuously),
$0.33-0.35$ (continuously); AO-3: $\psi=0.5-0.56$.
The time distribution of these observations is displayed in Fig.\ref{obsvstime}.
The $T_3$ moments are indicated
by squares and the closest orbital minima -- by triangles.

\section{Precessional variability}

\begin{figure}[ht]
\epsfig{file=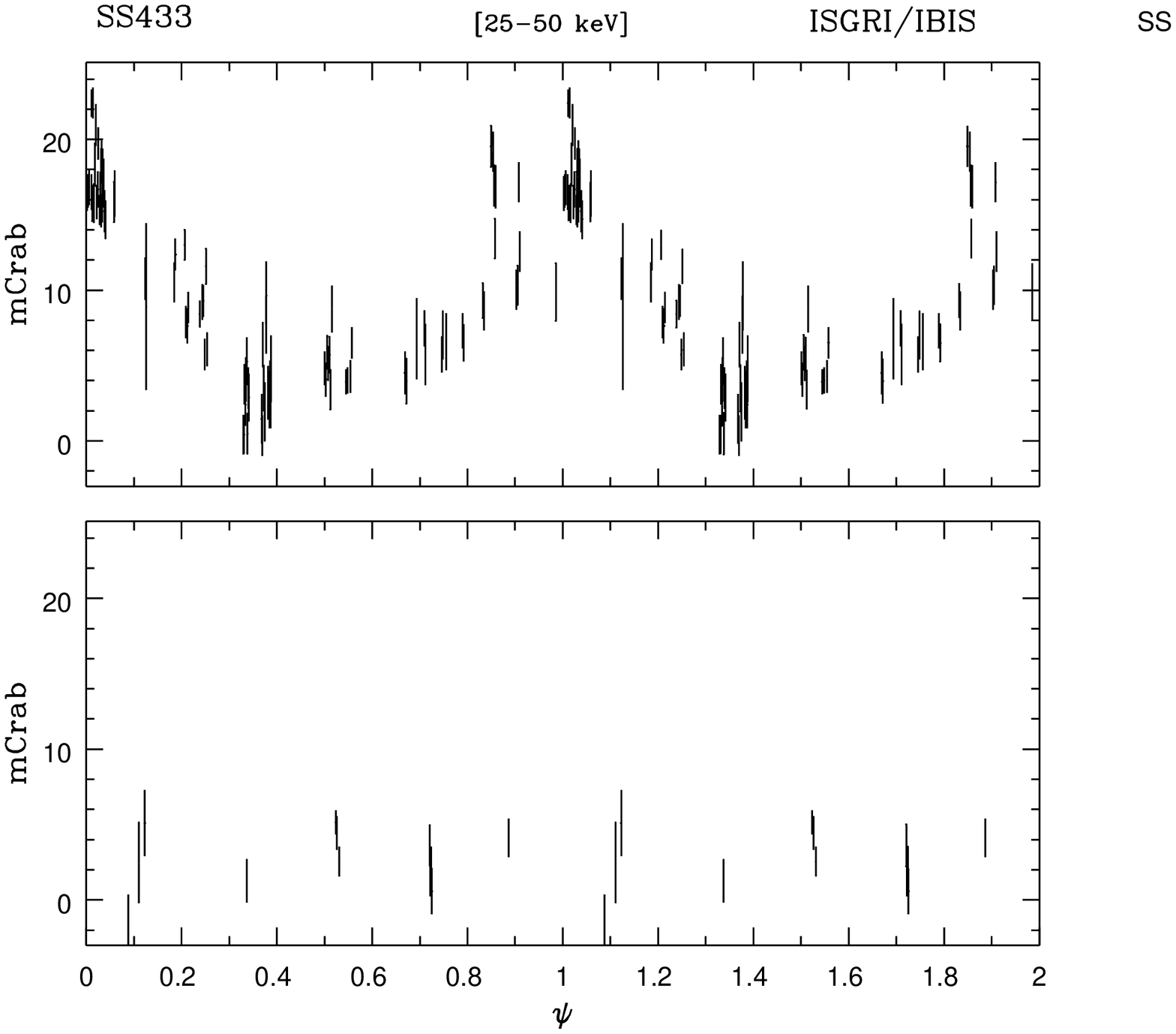,width=\linewidth}
\caption{Precessional light curves of \SS }\label{lcurve:prec}
\end{figure}

The $25-50$~keV light curve composed from all the available  ISGRI/IBIS 
observations of \SS folded with the precessional period $P_{prec}=162^d.5$ 
is displayed in Fig.~\ref{lcurve:prec}. 

The face-on disk position ($T_3$) defined as the 
maximim of the optical emission lines separation is taken as the moment of zero 
precession phase.  The upper and lower panels of Fig.~\ref{lcurve:prec} display
the X-ray flux (mCrabs) out of eclipse of the X-ray source by the normal star
and in the middle of the eclipse,
respectively.

Fig.~\ref{lcurve:prec} shows that the flux out of eclipse is varies from 
$3$~mCrab (crossover phase) to $18$~mCrab ($T_3$ phase). The 
flux in the middle  of eclipse (primary minimum) has no significant 
variablity, staying at average near $2.9$~mCrab.

\section{Orbital eclipse variability}

\begin{figure}[ht]
\epsfig{file=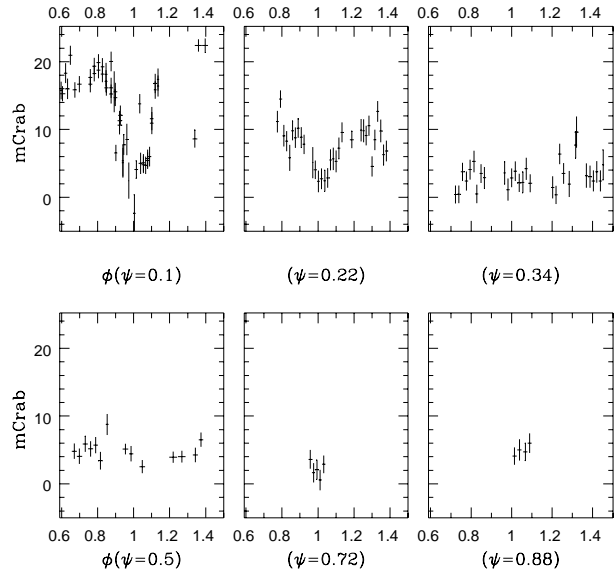,width=\linewidth}
\caption{Orbital light curves of \SS in different phases $\psi$ of precession.}\label{lcurve:orb}
\end{figure}

Fig.~\ref{lcurve:orb} displays the shape of the eclipses of \SS X-ray source by the normal component 
in the energy band $25-50$~keV observed in different phases $\psi$ of the period of precession.
The eclipse at $\psi=0.1$ is obtained by folding of all the available light curves
in the phases $\psi=0.95-0.15$ (JD 2452770-2453270).

The eclipse depth is maximal at $\psi=0.1$ and decreases with the phase of precession.
This seen clearly for individual eclipse light curves for $\psi=0.22,0.34$\ (first crossover), 
$\psi=0.5,0.72$ (some later than the second crossover), and $0.88$ (see Fig.~\ref{lcurve:orb}). 
The dotted line at the figure indicate the moments of the primary minimum.

The good correspondence between X-ray minima with the optical ephemeris for the precession phases 
close to $T_3$ as well as the delay of the X-ray light curve after the optical one at $\psi\sim0.5$
should be specially noted.

\begin{table*} 
\caption{Parameters of the SS433 spectrum for CompPS model}\label{sp_tab}
\begin{center}
\begin{tabular}{lccccc}
\hline
Model & $kT$ (keV)  & $\tau_y$ & $H/R$ &  Normalization & Reduced $\chi^2$\\ 
\hline
\hline
hemisphere & $24.9\pm 7.7$ & $1.8\pm0.6$ & -- & $13.3\pm 3.8$ & 1.08
(5 d.o.f.)\\
\hline
slab  & $23.2\pm 6$ & $2.8\pm 0.6$ & 0.5 & $12.7\pm 4.5$ & 1.06
(5 d.o.f.)\\  
\hline
\end{tabular}
\end{center}
\end{table*}  

\section{Eclipse in the phase $\psi=0.1$}

\begin{figure}[ht]
\epsfig{file=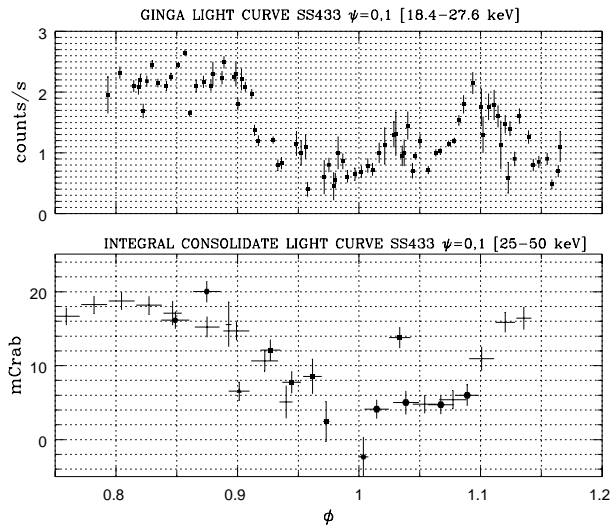,width=\linewidth}
\caption{The eclipse shape at $\psi=0.1$ }\label{lcurve:01}
\end{figure}

\begin{figure}[ht]
\centerline{
\epsfig{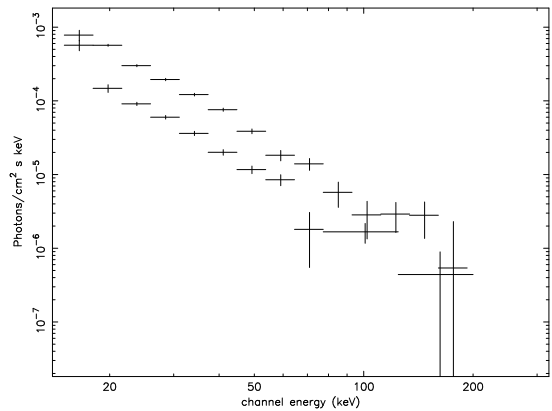} 
}
\caption{The IBIS/ISGRI $10-200$~keV high and low state \SS spectra (upper and lower points, respectively).}
\label{spectrum2}
\end{figure}

Fig.~\ref{lcurve:01} displays 
the eclipse light curves in the band $25-50$~keV obtained by \INT in comparison with the 
$18.4-27.6$~keV light curves obtained by Ginga (\cite{GingaObs}), denoted below as hard and soft bands.

Evidently, the eclipse width in the hard band is greater than in the soft one. This seems
unexpected because such  behaviour is inverse to that in regular X-ray binaries 
where the X-ray eclipse width decreases in harder band. 

The ascending part of the eclipse in the Ginga data is disturbed, probably, by the light absorption
by the accretion flows.
Some distortions at the same phases can be noticed also in the hard band data. 

The phase of the excess point inside the eclipse is coincident with the
local maximum on the Ginga light curve.

To avoid the influence of the accretion flows, we analyze only the descending part of the eclipse,
which can be considered as corresponding to the pure geometrical eclipse by the normal component body.

\section{Spectral analysis} 

\begin{figure}[ht]
\centerline{
\epsfig{file=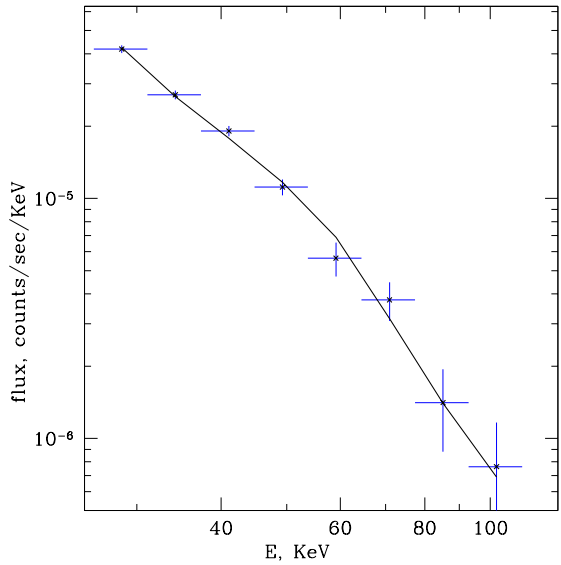,width=1.0\linewidth} 
}
\caption{The IBIS/ISGRI  $10-200$~keV  spectrum of \SS433  and its best fit
by thermal comptonization model CompPS with slab geometry (solid line).}
\label{spectra}
\end{figure}

For spectral analysis, we separated all available observations into 
two parts: below the mean X-ray flux (around 10 mCrab in the 20-50 keV energy range) 
and above it. 
The spectra where obtained using the software developed by \cite{Revn04} (IKI RAS).
No significant 
difference in the spectral shape was found (see Fig.~\ref{spectrum2}). This suggests
a fairly broad region emitting in hard X-rays with size 
compared to that of the accretion disk, since (apart from orbital eclipses) 
the observed flux varies due to precession of the disk. 
The spectral shape of this region in the 20-100 keV range 
can be fitted by 
thermal comptonization model (CompPS, \cite{Pout96}). 
Equally good fits were found for two cloud geometries: (1) a hemisphere
with temperature $kT=25\pm 8$ keV and optical thickness for electron
scattering $\tau_y=1.8\pm 0.6$, and (2) a
slab with $kT=23\pm 6$ keV and $\tau_y=2.8\pm 0.6$ (see 
Fig. \ref{spectra} and Table \ref{sp_tab}). In both cases the inclination
angle is assumed to be 60 degrees. 
A blackbody component with $kT_{bb}=1.8$ keV is added to slightly 
improve the fits.

\begin{figure}[ht]
\centerline{
\epsfig{file=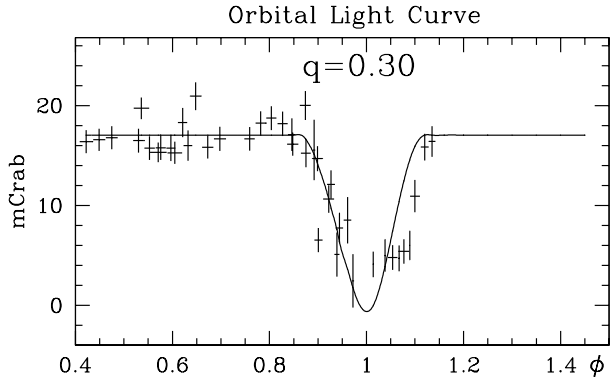,width=1.0\linewidth} 
}
\caption{Eclipse fit for the mass ratio $q=0.3$.}
\label{model:orb}
\end{figure}

This analysis suggests that in addition to thermal X-ray emission from jets 
(which dominates at smaller energies in the standard X-ray band, 
see \cite{Fil06} and references therein for earlier works), 
a broad hot corona emitting in hard X-ray exists in SS433 around jets, 
first discovered by \INT (Cherepashchuk et al \cite{Cher03}, \cite{Cher05}). From the
joint analysis of X-ray orbital eclipses and precessional variability 
presented in this paper it
follows that the size of the corona must be comparable with that of
the accretion disk ($10^{11}-10^{12}$ cm). For the physical parameters
obtained from spectral fits the electron denisty around $10^{12}$ cm$^{-3}$
can be derived. We note that such a denisty is indeed expected in the wind 
outflowing from a supercritical accretion disk with $\dot M\sim 10^{-4}$
M$_\odot$/yr and $v\sim 3000$ km/s at distances $\sim 10^{12}$ cm from 
the center where the photosphere should be formed. The heating 
of such a corona up to 25 keV  can be due to violent interaction 
of mildly relativistic jets with the surrounding 
dense wind (\cite{Beg06}).

\section{Analysis of orbital and precessional variability}

We carried out the joint analysis of orbital eclipses (ingress only) at the
precessional phase $\psi_{prec}\sim 0.1$ and of the precessional light curve
itself.

For the analysis of X-ray variability we used a geometrical model of SS433
applied earlier to the analises of Ginga data \cite{Ant92} and
the \INT X-ray light curve \cite{Cher05}. A close binary
system consists of an opaque ``normal'' star shaped by a Roche equipotential
surface and a relativistic object surrounded by an optically and
geometrically thick ``accretion disk''. The size of the ``normal'' star is
determined by the sizes of the critical Roche lobe for the mass ratio
$q=M_x/M_v$ (here $M_x$ is the mass of the relativistic object).

The ``accretion disk'' includes the disk itself and an extended photosphere
formed by the outflowing wind. The disk is inclined with respect to the
orbital plane by the angle $20.30^o$. The opaque disk body with a central
cone-like region (funnel) is described by the radius $r_d$ and the cone
opening angle $\omega$. The  accretion disk radius is limited
by the distance $r_d$ from its center to the inner Lagrangian point. The
relativistic object is surrounded by a transparent homogeneously emitting
spheroid with the visible radius $r_j$ and height $b_j$, which could be
interpreted as a ``corona'' or a ``thick jet'' (without any relativistic motion).

Only the hot ``corona'' is assumed to emit hard X-ray flux, while the star and
the opaque disk eclipse it during orbital and precessional motion.
Physically, such a ``corona'' could be thought of as hot low dense plasma
filling the funnel around the relativistic jets.

During precessional motion the inclination of the disk with respect to the
observer changes, causing different conditions of the ``corona'' visibility.
Observed precessional variability can thus be used to obtain a ``vertical''
scan of the emitting structure, restricting the height $b_j$ and the cone
opening angle $\omega$. The orbital variability (eclipses) allow one to scan
the emitting structure ``horizontally'', rectricting the possible values of
$r_d$, $r_j$, $\omega$ and the mass ratio $q$.

The results of the joint analysis of the orbital and precessional
variabilities are the following.

\begin{figure*}[ht]
\centerline{
\epsfig{file=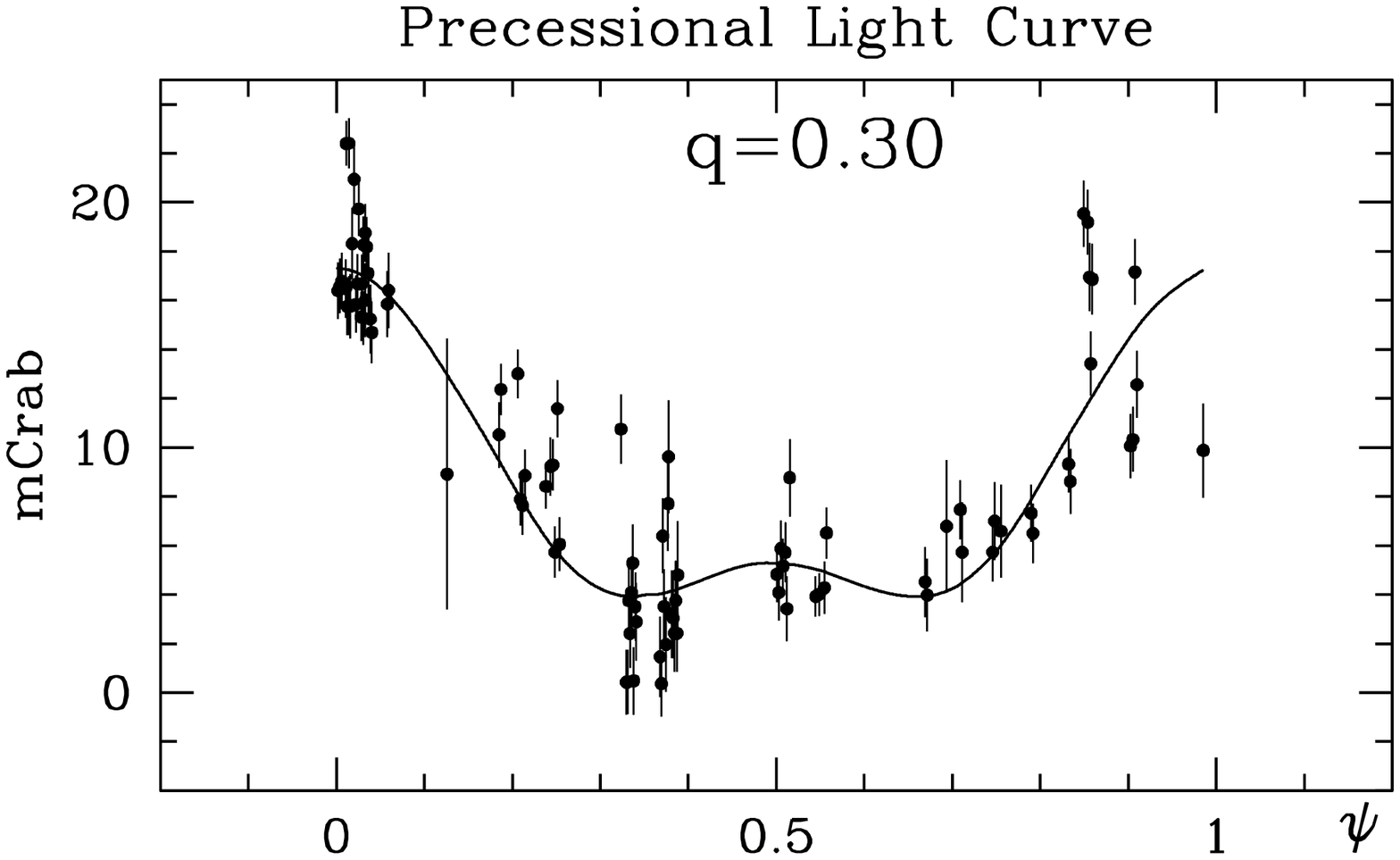,width=0.5\linewidth} 
\epsfig{file=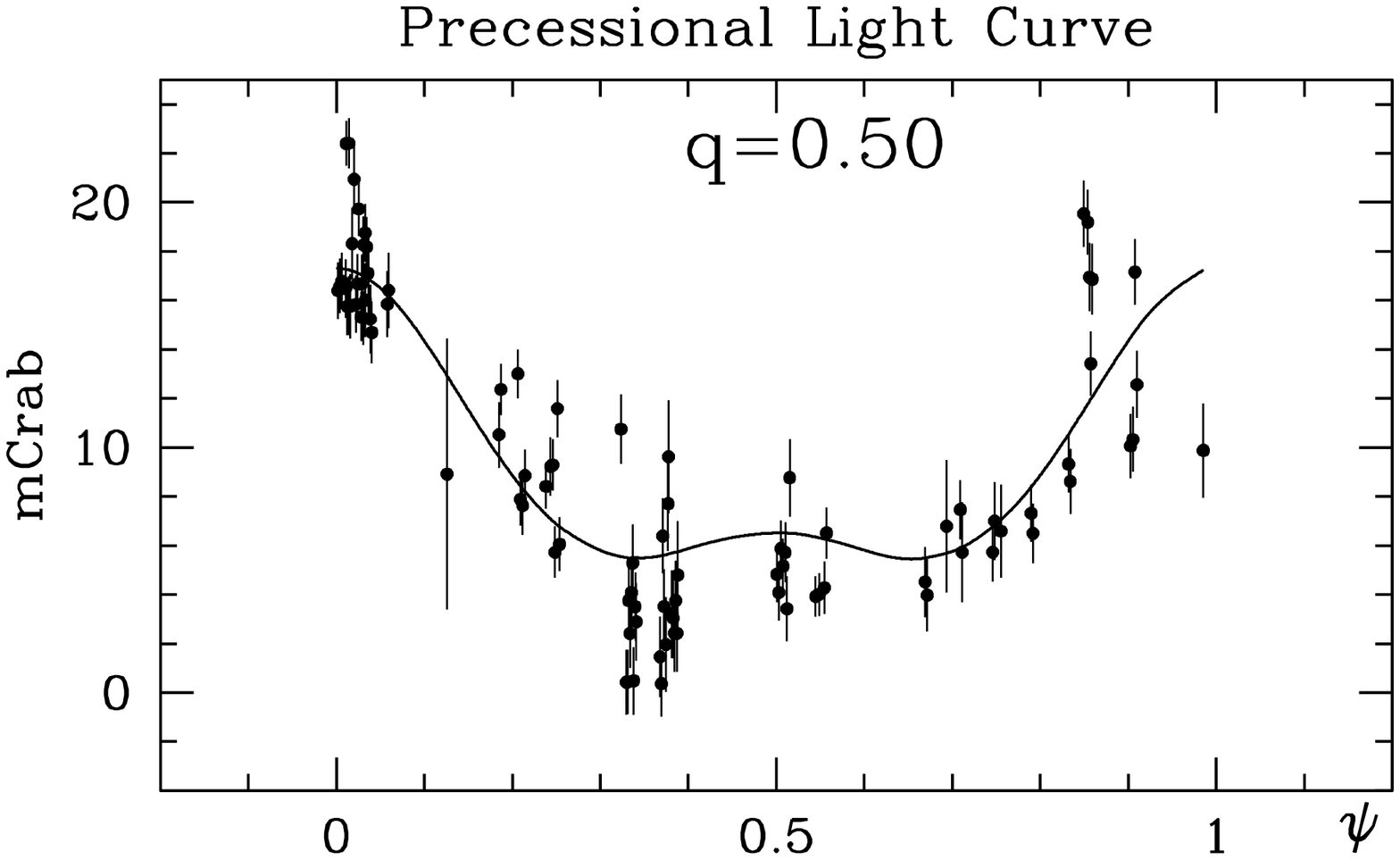,width=0.5\linewidth} 
}
\caption{Precessional light curve fits for $q=0.3$ and $0.5$ (left and right panels, respectively).}
\label{model:prec}
\end{figure*}

From the analysis we conclude that in the framework of our geometrical model
of the hard X-ray emitting region in SS433 the eclipse shape (ingress) and
the precessional amplitude are best reproduced by a wide oblate ``corona''
above an optically thick accretion disk.

At any $q$, the best fits for the observed wide orbital eclipse
($\psi_{prec}\sim 0.1$) are obtained for large visible radii of the hot
corona $r_j$ just slightly smaller than the disk radius $r_d$
($r_j/r_d=0.7-0.9$).  The height of the ``corona'' $b_j$ is about $0.3-0.6$
of $r_d$ and the cone opening angle $\omega=70-80^o$.

The precessional light curve taken alone does not restrict the binary mass
ratio $q$: equally good fits are found for $q$ varying from $0.05$
to $1.0$.

Analysis of the orbital X-ray eclipse ingress only (observed at
$\psi_{prec}\sim 0.1$) shows that in the range of $q=0.1-0.6$ the fit
accuracy
varies by some $10\%$ only.

Simultaneous analysis of both precessional and orbital variability of
SS433 allows one to constrain $q$ to the range $0.3-0.5$.
The fit is illustrated in Fig.~\ref{model:orb} and Fig.~\ref{model:prec}.


\begin{figure}[ht]
\epsfig{file=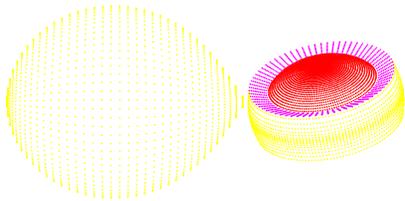,width=0.8\linewidth}
\caption{The graphical scheme of the emitting regions}
\label{model:schema}
\end{figure}

\section{Conclusions}

Most probably, \SS contains a blach hole. The detection of a hot extended corona around 
the supercritical accretion disk is a new result.

The future plans are connected with a high resolution spectroscopy of \SS at the face-on disk phase
(planned Subaru observations).
The new eclipse observations at $\psi=0-0.1$ in hard X-ray range 
would confirm the growth of the eclipse width with the 
photon energy, which would directly evidence the presence of an extended hot corona around the disk. 
Also the peculiarities of the ascending side of the eclipse are to be studied. These tasks are planned to 
be done by the \INT observations in May 2007.

\section*{Acknowledgments}

This work was partially supported by RFBR grants 06-02-16025 and 05-02-17489.


\begin{thebibliography}{}

\bibitem{Gor98} Goranskii, V.P., Esipov, V.F., Cherepashchuk, A.M., 1998, Astron. Reports, {\bf 42}, 209
\bibitem{Gies02} Gies, D.R., Huang, W., McSwain, M.V., 2002, ApJ {\bf 578}, L67
\bibitem{Hillwig04} Hillwig, T.C., Gies, D.R., Huang, W., et al 2004, ApJ {\bf 615}, 422 
\bibitem{Cher04} Cherepashchuk, A.M.Sunyaev, R.A., Fabrika, S.N. et al, 2004 
Proceedings of the 5th \INT Workshop on the \INT Universe (ESA SP-552). 
16-20 February 2004, Munich, Germany. Scientific Editors: V. Sch\"onfelder, G. Lichti \& C. Winkler, p.207
\bibitem{Cher05} Cherepashchuk, A.M., Sunyaev, R.A., Fabrika, S.N., Postnov, K.
A., Molkov, S. V., et al, \& Aph. 2005, {\bf 437}, 561
\bibitem{Barnes05} Barnes, A.D., Casares, J., Charles, P.A., Clark, J.S., Cornelisse, R.,
 Knigge, C., Steeghs, D., 2005, MNRAS, (astro-ph/0510448)
\bibitem{Fab04} Fabrika, S.N., 2004, Astrophys. Space Phys. Rev. {\bf 12}, 1
\bibitem{Cher03} Cherepashchuk, A.M., Sunyaev, R.A., Seifina, E.V., Panchenko, I.
E., Molkov, S. V., Postnov, K. A. 2003, Astron. \& Aph. {\bf 411}, L441
\bibitem{Cour03} Courvoisier T.J.-L., Beckmann V., Bourban G., et al. 2003, Astron. \& Aph., {\bf 411}, L53
\bibitem{GingaObs} Kawai N. and Matsuoka M., 1989, Publ. Astron. Soc. Japan, {\bf 41},491
\bibitem{Revn04} Revnivtsev M.G., Sunyaev R.A., Varshalovich D.A et al., PaZh, 2004 {\bf 30}, 430
\bibitem{Pout96} J. Poutanen, R. Svensson, 2006, ApJ, {\bf 470}, 249
\bibitem{Fil06} Filippova, E., Revnivtsev, M., Fabrika., S., Postnov, K., Seifina, E. 2006, Astron. \& Aph., in press. astro-ph/0609367
\bibitem{Beg06} Begelman, M.C.,  King, A.R.,  Pringle, J.E. 2006, MNRAS, {\bf 370}, 399
\bibitem{Ant92} Antokhina, E.A., Seifina, E.V., Cherepashchuk, A.M., 1992, Astron. Zh, {\bf 69}, 282


\end{thebibliography}
\end{document}